\documentclass[preprint,12pt]{elsarticle}
\usepackage{graphicx}
\usepackage[margin=1.0in]{geometry}
\usepackage{color, colortbl}
\usepackage{hyperref}
\usepackage{float}
\usepackage{lineno}
\usepackage{tabularx} 
\usepackage{array}    
\usepackage{subcaption}
\usepackage{listings}
\usepackage{orcidlink}

\definecolor{LightCyan}{rgb}{0.88,1,1}
\definecolor{LightRose}{rgb}{1,0.88,0.88}
\definecolor{LightGreen}{rgb}{0.88,1,0.88}

\definecolor{codegreen}{rgb}{0,0.6,0}
\definecolor{codegray}{rgb}{0.5,0.5,0.5}
\definecolor{codepurple}{rgb}{0.58,0,0.82}
\definecolor{backcolour}{rgb}{0.95,0.95,0.92}

\lstdefinestyle{mystyle}{
    backgroundcolor=\color{backcolour},   
    commentstyle=\color{codegreen},
    keywordstyle=\color{magenta},
    numberstyle=\tiny\color{codegray},
    stringstyle=\color{codepurple},
    basicstyle=\ttfamily\footnotesize,
    breakatwhitespace=false,         
    breaklines=true,                 
    captionpos=b,                    
    keepspaces=true,                 
    numbers=left,                    
    numbersep=5pt,                  
    showspaces=false,                
    showstringspaces=false,
    showtabs=false,                  
    tabsize=2
}

\title{High-Performance Data Format for Scientific Data Storage and Analysis}

\author[1]{Gagik Gavalian \orcidlink{0000-0002-6738-5457}}
\address[1]{Thomas Jefferson National Accelerator Facility, Newport News, VA, USA}

\begin{document}

\begin{abstract}
In this article, we present the High-Performance Output (HiPO) data format developed at Jefferson Laboratory for 
storing and analyzing data from Nuclear Physics experiments. The format was designed to efficiently store large
amounts of experimental data, utilizing modern fast compression algorithms. The purpose of this development was 
to provide organized data in the output, facilitating access to relevant information within the large data files. The HiPO
data format has features that are suited for storing raw detector data, reconstruction data, and the final physics analysis 
data efficiently, eliminating the need to do data conversions through the lifecycle of experimental data. The HiPO data format
is implemented in C++ and JAVA, and provides bindings to FORTRAN, Python, and Julia, providing users with the choice of data 
analysis frameworks to use.
In this paper, we will present the general design and functionalities of the HiPO library and compare the performance of the library with
 more established data formats used in data analysis in High Energy and Nuclear Physics (such as ROOT and Parquete). In columnar data analysis, HiPO surpasses established data formats in performance and can be effectively applied to data analysis in other scientific fields.

\end{abstract}
\maketitle

\section{Introduction}
\label{section-introduction}

Over the decades, high-energy experiments, such as those conducted at particle accelerators like CERN’s Large Hadron Collider (LHC), have seen an exponential increase in data production. Early particle physics experiments in the mid-20th century relied on photographic plates and bubble chambers, which generated manageable amounts of data for manual analysis. As technology advanced, detectors became more sophisticated, capturing finer details of particle collisions across multiple dimensions, from spatial tracks to energy signatures. This evolution enabled scientists to probe deeper into the structure of matter and the fundamental forces of nature but came with an enormous uptick in data volume.

Modern experiments produce petabytes of data annually, thanks to the use of advanced digital detectors and high-frequency collision rates. For instance, the LHC can generate up to one billion proton-proton collisions per second during its operations. Each collision results in complex, high-dimensional data that must be recorded, processed, and analyzed. However, the vast majority of these collisions are routine and unremarkable, reflecting well-known physics processes. Only a tiny fraction contains the rare and novel events that could reveal new particles or phenomena, such as the discovery of the Higgs boson in 2012.

The challenge lies in efficiently processing this deluge of data to identify and retain only the relevant information while discarding the rest. This is achieved through a multi-layered system of data selection. First, hardware-based triggers operate in real-time to reduce data rates by orders of magnitude, selecting events based on basic characteristics like energy thresholds. Then, software-based algorithms provide more refined filtering by analyzing the remaining data for specific patterns or anomalies. Despite these strategies, the volume of "useful" data still reaches hundreds of terabytes, requiring vast computing resources and distributed networks to store and process the information.

Another key challenge is ensuring that the data reduction process does not inadvertently discard valuable signals. Designing selection algorithms requires balancing sensitivity to rare events with the need to filter out noise effectively. Advances in machine learning have become increasingly important in addressing this problem. Sophisticated models can analyze high-dimensional data for subtle correlations and anomalies, improving the ability to identify rare events. However, the computational demands of such models add another layer of complexity, requiring extensive computing power, energy, and expertise.

The rise of big data in high-energy physics not only highlights the field’s technological achievements but also underscores the growing need for innovative solutions in data management and analysis. Collaboration between physicists, computer scientists, and engineers will remain critical in addressing these challenges and unlocking the secrets hidden in the vast streams of experimental data.

\section{Motivation}
\label{section-motivation}
Physics experiment data comprises "events," each representing the information captured from a single interaction between an incident beam particle and the target. These events are processed individually to identify particles and reconstruct their trajectories using data from various detector components. This process creates a complete physics event, which serves as the foundation for advanced physics analyses. Each event encapsulates responses from all detector components, organized into distinct data structures. Before reaching the physics analysis stage, the data undergoes several transformations to prepare and refine it for further study.
\begin{itemize}
\item {\bf Data Acquisition:} The system records raw data from all detector components for each interaction instance. This raw data typically includes time measurements and accumulated charges for each detector system component.
\item {\bf Data Transformation:} In the next phase, the raw data is converted into calibrated values with physical units, such as time in milliseconds and energy in electron volts (eV).
\item {\bf Reconstruction:} A reconstruction program analyzes data from individual detectors to identify related signals, then integrates signals across various detectors to identify particles in each event. The output includes tables detailing particle information and their responses in each detector component, aiding in particle species identification.
\item {\bf Physics Analysis:} Tailored selection algorithms are applied to identify specific physics reactions within each event. Physics observables are calculated based on the detected particles, and the final output is organized into a columnar table for high-level physics analysis.
\end{itemize}

In traditional CLAS~\cite{CLAS:2003umf} experiments, different data formats were used at each stage of the data lifecycle. This approach added unnecessary complexity, requiring the support of multiple file formats and the maintenance of various conversion tools. Additionally, users developed numerous data selection and filtering tools tailored to these formats, all demanding regular maintenance. Initially, a similar strategy was proposed for the CLAS12~\cite{Burkert:2020akg} experiment during its early software development phases.

To address these limitations, the High-Performance Output (HiPO)~\cite{hipo5p0:2025jk} data format was developed specifically for the CLAS12 experiment. HiPO is designed to efficiently handle all stages of experimental data processing, from reconstruction workflows to final columnar data analysis. It offers language bindings for various programming languages used within the collaboration, including C++, FORTRAN, Python, Java, and Julia, ensuring broad compatibility and streamlined workflows.

To meet the demands of all data processing workflows, several key requirements were established for the data format:
\begin{itemize}
\item {\bf Serializable:} The CLAS12 reconstruction workflow employs a Service-Oriented Architecture (SOA) operating on a heterogeneous platform with message passing. HiPO was designed to be easily serializable, allowing event data to be transmitted efficiently to individual reconstruction services.
\item {\bf Compression Efficiency:} To reduce storage demands, HiPO incorporates a compression algorithm that balances speed and compression ratio. High compression speed is critical to managing the large data volumes generated in high-rate nuclear physics experiments, ensuring high data throughput.
\item {\bf  Random Access Capability:} HiPO supports random access to specific data collections within a file. This feature is essential for debugging, selective data writing, and multi-threaded applications that process data chunks asynchronously.
\item {\bf Data Grouping Functionality:} HiPO enables the grouping of related datasets, allowing efficient tagging or marking of different datasets. This functionality facilitates the targeted retrieval of specific groups without needing to process the entire dataset.
\end{itemize}
The CLAS12 data reconstruction framework is implemented in Java, making Java the primary development platform for the HiPO library. A parallel C++ library is also developed, though it sometimes lags behind Java in terms of features, with new functionalities gradually ported to C++. Most of the examples provided in this document are written in Java, with equivalent C++ examples available in the HiPO repository. Experimental bindings for Python and Julia exist but are not actively developed due to limited usage by collaborators.

The subsequent chapters will delve deeper into the features of the HiPO data format, complemented by illustrative examples.

\section{Format Description}
\label{section-format-description}

Physics experiment data is organized into discrete units called "events," where each event represents the data associated with a single physics interaction recorded by the detector. These events are gathered over a specific timeframe and stored sequentially within a file. A HiPO file, tailored for this purpose, is composed of the following components:

\begin{figure}[h!]
  \begin{center}
    \includegraphics[width=0.85\textwidth]{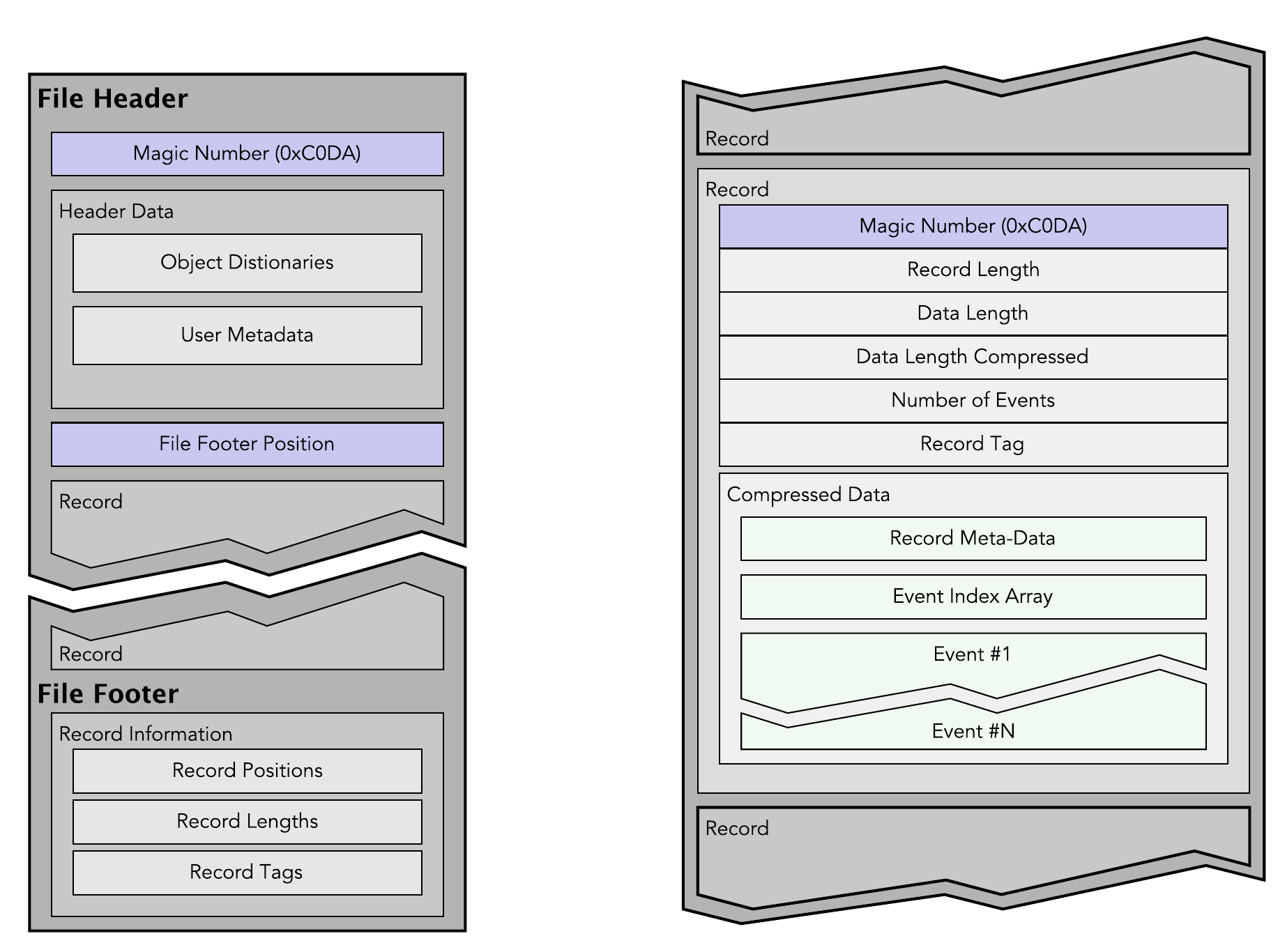}
 \end{center}
  \caption{Schematic view of the HiPO file structure. The figure shows the conceptual design of the file structure. It's not an exact layout of the data structures of the headers. The detailed documentation is published on the GitHub repository with the source code.}
 \label{schema:file}
\end{figure}

\begin{itemize}
\item {\bf File Header:} This section includes essential information such as the file version, file length (used for consistency checks), the location of the file footer, and other relevant parameters.
\item {\bf User Header:} This part contains metadata that describes the content of the file any other user provided metadata. In dictionary-driven formats, it includes descriptors (schemas for tabular data) of the data stored within the file.
\item {\bf Data Records:} These are the compressed, grouped data from individual events. The size of data records is configurable at the time of file creation. Each record is assigned a user-defined identifier, known as a tag, to group similar events together.
\item {\bf File Footer:} The footer contains metadata about each data record in the file, including its position, size, and tag.
The general structure of a HiPO file is illustrated in Figure~\ref{schema:file}.
\end{itemize}

The schematic view of the file structure is shown in Figure~\ref{schema:file}. 

The file header serves as a descriptor for the file, containing essential information such as the file version and a "magic word" that identifies the file format. It also specifies the data's endianness and the location of the file footer. Additionally, the header includes a "User Data" block, which contains dictionaries for the data objects stored in the file and user-defined metadata. This metadata, typically provided during the file initialization stage, is composed of key-value pairs that describe the file or include any parameters the creator wishes to add. To create a basic file with a specified user header, refer to the code in Listing~\ref{lst:open_file}.

\begin{lstlisting}[language=java, caption=Java example for writing a file with meta-data. The reader object opens the file and retrieves the stored mata data in a form of a key/value map., label=lst:open_file]
// -- open file with user-specified meta data
HipoWriter w = new HipoWriter(); // create the object
w.addConfig("date","created on 01/27/2021");
w.addConfig("author","John Pierce");
w.addConfig("type","Physics Events from A-Detector")
w.open("my_new_file.h5");
w.close();
// -- read metadata from the file
HipoReader r = new HipoReader("my_new_file.h5");
Map<String,String> userInfo = r.getUserConfigurations();
\end{lstlisting}

\subsection{Records}

Records store event data sequentially, with each record containing a header that provides key information. This includes the number of events stored, an array of indices pointing to each event within the buffer, and the length of the data payload both before and after compression. The maximum size of records in a file is configured at the time of file creation, with a default size of $8~{MB}$. 

The record header also includes a unique identifier, set to "0" by default. Events are sequentially added to the record until the size limit is reached. Once the limit is met, the record is compressed, written to disk with its header attached, and then cleared to receive new events. This cycle repeats until the file is closed.

When the file is closed, a file footer is created, which contains the positions and identifiers of all records. Additionally, the file header is updated to include the location of the footer, enabling fast access when the file is reopened for reading.

\subsection{Events}

As previously mentioned, a HiPO file consists of a series of events. An event represents a single unit containing a collection of interrelated data objects. The types of data that can be stored in an event include arrays of primitive types and tables. Each object within an event is assigned two identification numbers, referred to as "group" (16 bits) and "item" (8 bits). These identifiers facilitate the retrieval of objects during reading and can be used creatively to organize related data. 

Primitive types are represented as simple arrays, with their headers fully describing the data structure. As a result, they do not require additional dictionaries to be interpreted.

\subsubsection{Primitive types}

Primitive types are arrays of numbers or strings, where all elements within the array share the same data type (e.g., byte, short, integer, float, double, long, or string). Objects, referred to as ``Nodes``, are created from these arrays using user-defined identifiers. Listing~\ref{lst:write_arrays} demonstrates how to create primitive arrays and write them into an event.

\begin{lstlisting}[language=java, caption=Java example to create and write primitive types into an event, label=lst:write_arrays]
// Writing arrays into an Event
Event event = new Event(2048); // creta event with max size 2 kB
float[]  df = new float[]{1.0,2.0,3.0,4.0};
short[]  ds = new short[]{3,5,8,13,21,34,55};
Node   nf = new Node(12,1,df);
Node   ns = new Node(12,2,ds);
Node data = new Node(12,3,"Event recorded at 12:52:33"); 
event.write(nf);
event.write(ns);
event.write(data);
\end{lstlisting}

It is important to note that the initial size of an event does not limit the ability to add objects exceeding the allocated event size. As new objects are added, the event buffer dynamically adjusts its size to accommodate them. However, for optimal performance, it is recommended to set the initial size slightly larger than anticipated to minimize reallocations.

\subsubsection{Tables}

Tables are rectangular data structures organized into rows and columns. To parse the content of a table, a schema (descriptor) is required. This schema must be defined and declared before the file is opened for writing, as the file writer compiles a dictionary and stores all schemas in the file header. The data structure used to manage tables is called a ``Bank``, which is also assigned two unique identifiers (``group`` and ``item``). Listing~\ref{lst:write_bank} provides a straightforward example of declaring and writing a simple bank into a file.

\begin{lstlisting}[language=java, caption=Java example to create and write banks (tables) into an event, label=lst:write_bank]
// Writing banks to the event
SchemaBuilder b = new SchemaBuilder("data::clusters",12,1)
    .addEntry("type","B","cluster type") // B - type Byte
    .addEntry("n", "S", "cluster multiplicity") // S - Type Short
    .addEntry("x", "F", "x position") // F - type float
    .addEntry("y","F","y position") // F - type float
    .addEntry("z", "F", "z position"); // F - type Float
// -- create a cchema
  Schema schema = b.build();
// add schema to the file and open the file
  HipoWriter w = new HipoWriter();
  w.getSchemaFactory().addSchema(schema);
  w.addConfig("date","file created at 11:54:22 AM");
  w.addConfig("description","file contains clusters in calorimeter");
  w.open("clusters.h5");                
  Event event = new Event();
  Bank  b  = new Bank(schema,2); // create a table with 2 rows
  b.putByte("type", 0, (byte)   1); b.putByte("type", 1, (byte)   2);
  b.putShort("n"  , 0, (short) 13); b.putShort("n"  , 1, (short) 21);
  b.putFloat("x",0,0.1f); b.putFloat("x",1,0.2f);
  b.putFloat("y",0,1.1f); b.putFloat("y",1,1.2f);
  b.putFloat("z",0,2.1f); b.putFloat("z",1,2.2f);
  event.write(b);
  w.addEvent(event);
  w.close(); // close should be called to write the file footer
\end{lstlisting}

The methods for writing and reading values in a bank provide two interfaces: one uses the name of the variable, enhancing code readability, while the other uses the column index, offering better performance when required. For example, the same setter can be used as follows: { \tt b.putFloat("x", 0, 0.1f)} or {\tt b.putFloat(2, 0, 0.1f)}, where {\tt "x"} is the third column in the table (column indices start at 0). 

When writing data, it is important to use the correct type-specific setters to prevent values from exceeding the type boundaries. However, more generic getters can be used when reading data, such as {\tt getInt(entry, row)} for all integer types and {\tt getDouble(entry, row)} for all floating-point types.

\begin{lstlisting}[language=java, caption=Java example to read banks from the file, label=lst:read_bank]
HipoReader r = new HipoReader("clusters.h5");        
Bank[] banks = r.getBanks("data::clusters");
// to read more than on bank use: r.getBanks("a","b","c","d");
// If the bank is not present in the event, the returned object 
// will have getRows()==0, no error is generated.
while(r.nextEvent(banks)){
  System.out.printf("\%4d, \%5d, \%8.5f \%8.5f \%8.5f\n",
           banks[0].getInt("type", row),
           banks[0].getInt(1,row),
           banks[0].getFloat("x", row),
           banks[0].getFloat("y", row),
           banks[0].getFloat("z", row));
}
\end{lstlisting}

Listing~\ref{lst:read_bank} demonstrates how to read banks from a file and display their content on the screen. The {\tt getBanks(String... list)} method accepts a list of bank names to be read for each event, allowing multiple banks to be processed and analyzed simultaneously. For more advanced examples of reading events and querying their contents, refer to the examples provided in the repository.

The code in Listing~\ref{lst:read_bank} illustrates a straightforward procedure where banks are automatically read for each event. More complex examples can be found in the repository, showcasing scenarios where the `Event` object is read from the file and banks are individually extracted, modified, deleted, or augmented before being written back to the output file.

\section{Event Tagging}

The HiPO is designed to store data in units called events, which are collections of related data, typically organized by time. In nuclear physics experiments, an event corresponds to a single interaction between a beam particle and a target, capturing the resulting particles from that interaction. After data acquisition, post-processing identifies the number and properties of the particles produced, recording this information within the event for further physics analysis. Different interactions occur at varying frequencies in these experiments, and specific analysis programs focus on particular physics reactions. These programs do not analyze every event but instead require events with a certain number of reconstructed particles to proceed.

\begin{figure}[h!]
  \begin{center}
    \includegraphics[width=0.85\textwidth]{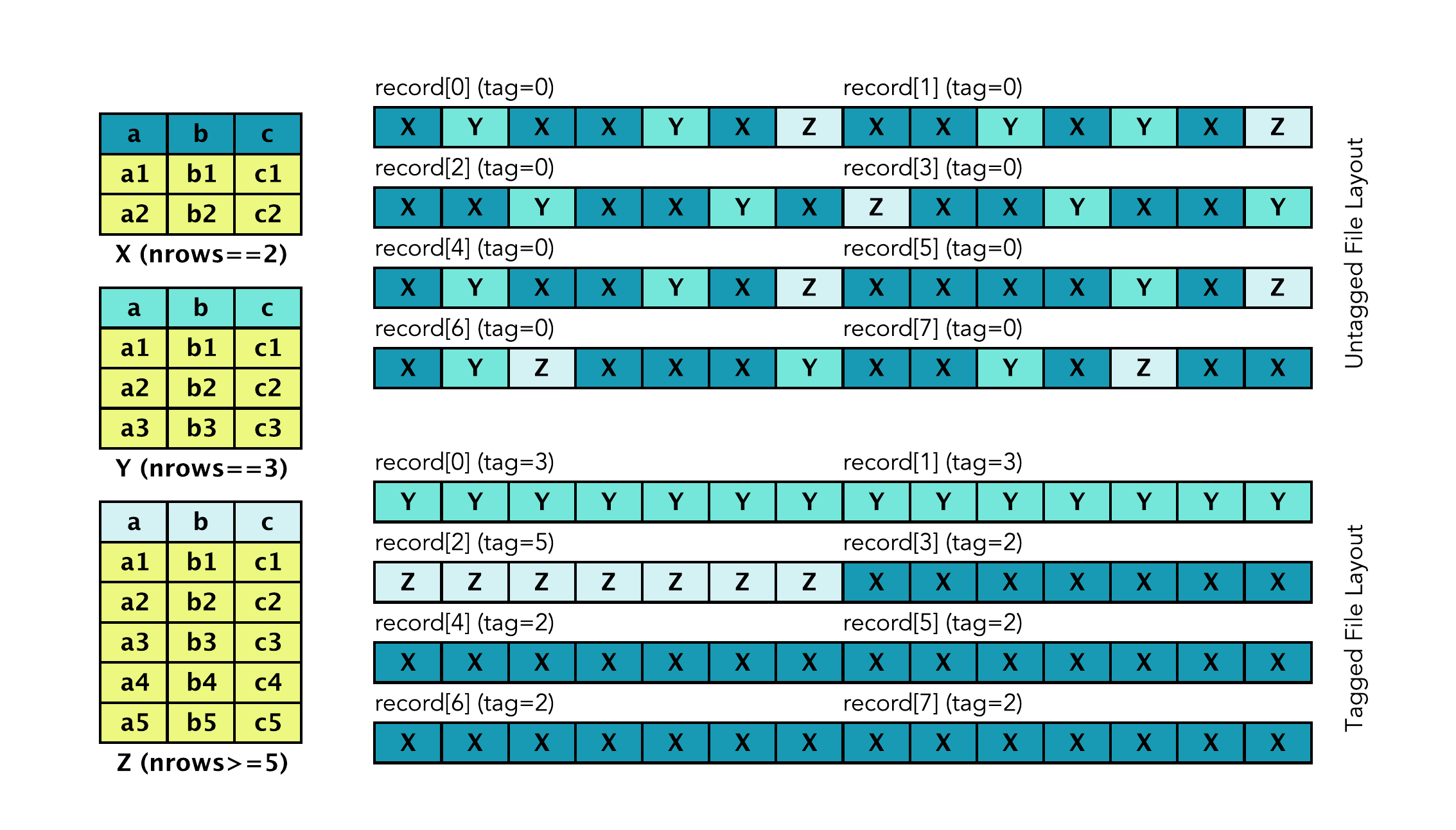}
 \end{center}
  \caption{Schematic view of the file output for tagged and untagged events. When using the tagged mode of writing a file, events are organized in the records 
  based on the tag of the event, which is assigned by the user depending on the needs.}
 \label{fig:event_frequency}
\end{figure}

Figure~\ref{fig:event_frequency} illustrates a schematic view of how data can be theoretically distributed within a file. Event types are represented by rows in a table listing the particles reconstructed by data processing software. For rare interactions, much of the data read is irrelevant to the analysis, leading to inefficient use of I/O resources. This inefficiency places unnecessary strain on shared file systems and computational resources.

To address this issue, the HiPO data format incorporates a tagging feature for organizing stored data. Each record in the file is assigned a unique tag identifier, and this information is stored in the file footer along with the record's positions and sizes. This approach enables events to be sorted into separate records during the file-writing process, grouping similar events together. Listing~\ref{lst:write_tagged} provides an example of tagging events based on the number of rows in a table that lists clusters, as described in the previous example.

\begin{lstlisting}[language=java, caption=Java example to create and write primitive types into an event, label=lst:write_tagged]
// Writing arrays into an Event
HipoReader r = new HipoReader("myfile.h5");
// using HipoWriter.create() transfers all the dictionary
// and the metadata to the write object
HipoWriter w = HipoWriter.create("taggedfile.h5",r);

Bank[] b = r.getBanks("data::clusters");
Event event = new Event();
while(r.next(event)){
  event.read(b);
  if(b[0].getRows()>=5) event.setTag(5);
  if(b[0].getRows()==2) event.setTag(2);
  if(b[0].getRows()==3) event.setTag(3);
  w.addEvent(event);
}
w.close();
\end{lstlisting}

The example code generates a file in which all events with the same number of rows are grouped together. Events with a specific number of rows in the "clusters" bank can be accessed directly, eliminating the overhead of scanning through the entire file. An example of reading events where the number of rows in the "clusters" bank is equal to 2 or greater than 4 is provided in Listing~\ref{lst:read_tagged}.

\begin{lstlisting}[language=java, caption=Java example to sort events in the output file depending on number of rows in the table, label=lst:read_tagged]
// Writing arrays into an Event
HipoReader r = new HipoReader();
r.setTags(2,5); // read only tag=2 and tag=5
r.open("taggedfile.h5");
Bank[] b = r.getBanks("data::clusters");
while(r.nextEvent(b)){
   b[0].show();// print bank content on the screen
}
\end{lstlisting}

In Listing~\ref{lst:write_tagged}, only the "data::clusters" bank containing either two rows or more than four rows is read. However, when an event is written to the output file, the entire event, including all banks and nodes, is recorded. The decision on how to partition the file, though, is based solely on the contents of the specified bank. This feature is utilized in CLAS12 to tag events based on their topology, such as the number and types of detected particles. This approach simplifies the process of reading only the relevant events from the data file for analysis.

\section{Columnar Data Storage}

The final stage of data analysis involves representing the data in a columnar format and retrieving subsets of data based on constraints applied to other columns. At this stage, users work with large datasets containing numerous columns and rows, where access to every column in the file is typically unnecessary. 

The Apache Parquet~\cite{PARKQUET:2020jk} format is widely used in data science for storing columnar data due to its compatibility with Python data frames and its efficiency in managing large datasets. Parquet is highly optimized for storing vast amounts of data and allows for fast access to specific columns, making it increasingly popular in High Energy and Nuclear Physics as Python gains traction as a preferred analysis environment. 

Historically, ROOT has been the primary tool for physics analysis over the past two decades. It provides robust data structures for storing columnar data and supports selective access to columns. Notably, ROOT also offers Python bindings, enabling integration with modern analysis workflows.

To expand the utility of HiPO files beyond data processing and into physics analysis, an experimental implementation of columnar data storage has been developed and tested. This implementation was evaluated against more established data formats to explore its potential in this domain.

\subsection{Design}

The HiPO data format offers flexibility for storing data in a columnar format, similar to Parquet and ROOT. By leveraging the feature of assigning tags to records, data can be organized column-wise, with each column assigned a unique tag and written into its own record. 

\begin{figure}[h!]
  \begin{center}
    \includegraphics[width=0.85\textwidth]{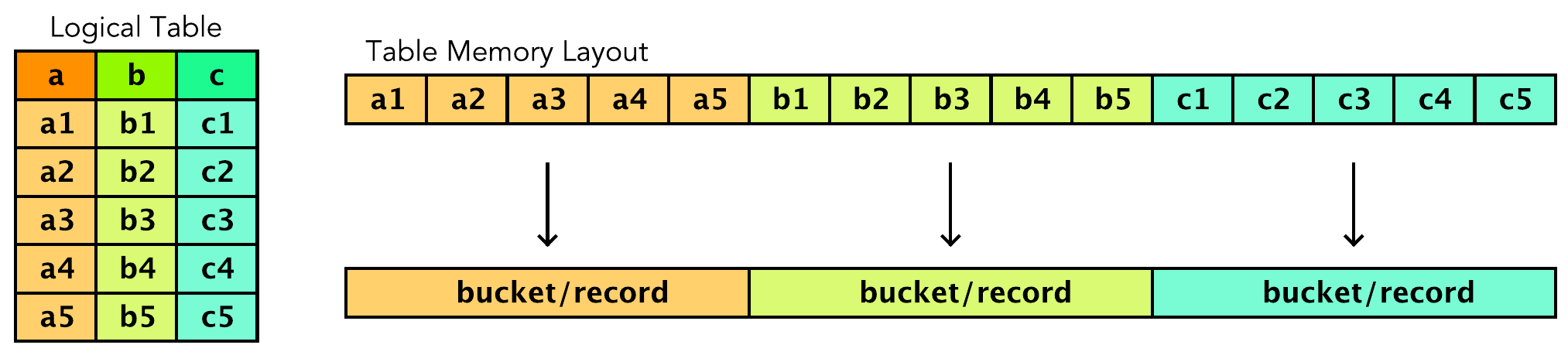}
 \end{center}
  \caption{Schematic view of arranging each event in the file from columnar table.}
 \label{fig:tuple_schema}
\end{figure}

This process is synchronized such that, after a predetermined number of rows, the data for each column is serialized and written as a separate record. A schematic representation of this data arrangement within a file is illustrated in Figure~\ref{fig:tuple_schema}.

When reading a file, the desired columns (referred to as branches) are specified, and only records (buckets) with matching tag numbers are read and deserialized. The user program can access only the declared branches, ensuring efficient data handling. 

\begin{lstlisting}[language=c++, caption=c++ example to read HiPO tuple file and fill a histogram., label=lst:read_tuple]
// open file and read-only specified branches
hipo::tuple tuple("tuple.h5", "c1:c2:c3:c4"); 
float data[4]; // declare a holder for the data to be read
twig::h1d h(120,-1.0,1.0); // declare a histogram
while(tuple.next(data)==true){
    h.fill(data[0]);
}
\end{lstlisting}

Files written in a columnar format must be read using the corresponding API to ensure proper synchronization of columns during the reading process. Below is an example code snippet demonstrating how to read specific columns from a file and populate a histogram.

The example code in Listing~\ref{lst:read_tuple} demonstrates how to open a file and read the branches named "c1" through "c4." It processes all rows in the file, passing them to the user code via a declared array. The values from the first branch ("c1") are then used to populate a histogram.

\subsection{Benchmarking}

For reading tests, we produced a synthetic data set consisting of 24 columns and 50 M rows and created HiPO (4.8 GB), ROOT (4.4 GB), and Parquete (5.1 GB) files, all three with LZ4 compression. The columns were filled with randomly generated numbers in the range $0..1$. 

\begin{figure}[h!]
  \begin{center}
    \includegraphics[width=0.85\textwidth]{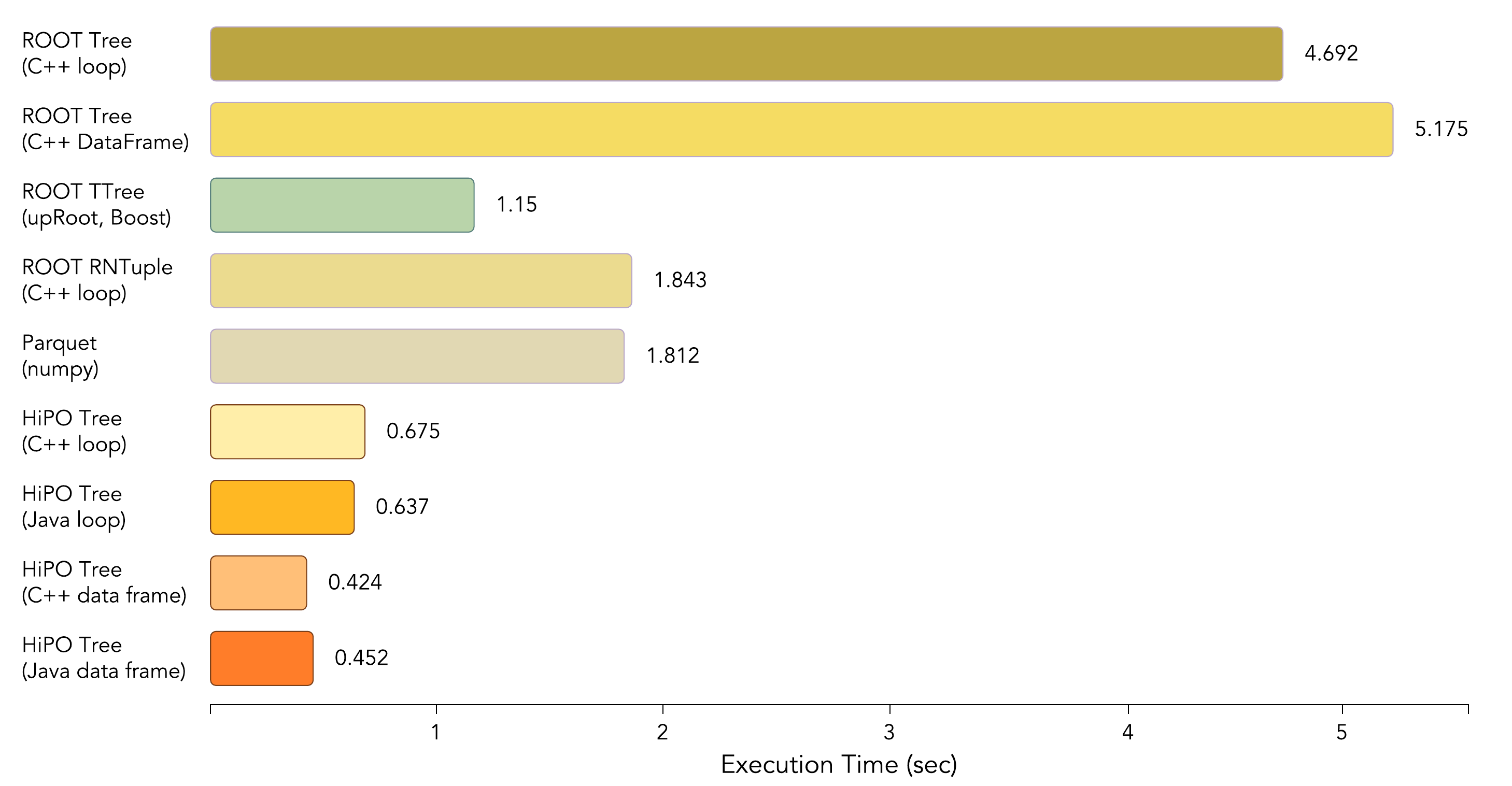}
 \end{center}
  \caption{Reading benchmark for different file formats reading the columnar data with 24 columns and 50 M rows (file size $\approx 4.5 GB$). Only four columns are read, and histograms are filled for the respective columns. }
 \label{read_benchmark}
\end{figure}

Reading tests were conducted by extracting four branches (out of 24) from the data file and populating a histogram for each branch, simulating a typical data exploration workflow. The tests measured the time taken for ten consecutive reads, with the average time of the last four reads used as the final result. The results are presented in Figure~\ref{read_benchmark}. 

As shown in the figure, the ROOT TTree exhibited the poorest performance. However, the RNTuple, the new data format standard in the upcoming ROOT 7 release, demonstrated significant improvements in data reading speeds, achieving performance levels closer to those of Parquet.

\begin{table}[h!]
\centering
\begin{tabular}{|p{7cm}|p{3.0cm}|p{4.0cm}|}
\hline
\textbf{Format (Library)} & File Size (GB)  &  Execution time (sec)  \\ \hline \hline
Tuple HiPO Java loop      & 4.8&0.637 \\ \hline
Tuple HiPO Java DataFrame &4.8& 0.452 \\ \hline
Tuple HiPO C++  loop      &4.8& 0.675 \\ \hline
Tuple HiPO C++  DataFrame &4.8& 0.424 \\ \hline
Tuple HiPO Julia  DataFrame &4.8& 0.487 \\ \hline
Tuple Parquete DataFrame  &5.1& 1.812 \\ \hline
TTree ROOT C++ loop       &4.4 & 4.692 \\ \hline
TTree ROOT C++ DataFrame  &4.4 & 5.175 \\ \hline
RNTuple ROOT C++ loop     &3.8& 2.624 \\ \hline
RNTuple ROOT C++ loop (CINT)    &3.8&  5.283 \\ \hline
Tuple HiPO Java loop (JShell)  &4.8&  0.672 \\ \hline
\end{tabular}
\caption{Reading benchmark for different file formats reading four columns from a file and filling respective histograms for each column. The file is generated with 24 columns and 50M rows filled with random numbers [0..1]. }
\label{tab:read_benchmark}
\end{table}

The numerical results of these tests are summarized in Table~\ref{tab:read_benchmark}. The HiPO data format demonstrates significantly faster reading speeds compared to Parquet and ROOT (RNTuple) when performing simple data reading in a loop (as illustrated in Listing~\ref{lst:read_tuple}), in both C++ and Java implementations. The "DataFrame" column in the table represents an experimental HiPO feature where histograms are passed directly to the tuple reader, utilizing bulk histogram filling from data buckets. This method achieves a $50\%$ improvement in histogram filling efficiency.

The table also includes preliminary benchmarks for the Julia port of the HiPO library (not shown in the graph), which exhibits performance comparable to compiled C++ code.

It's worth noting that much of exploratory data analysis is conducted in interactive environments using ROOT files, where variables are plotted with various selection combinations. Our tests revealed that RNTuple performance degrades significantly in interactive mode (by a factor of 2). In contrast, the HiPO Java library, when used in interactive **JShell**, maintains its performance, as reflected in Table~\ref{tab:read_benchmark}.

Finally, while HiPO's columnar storage feature is still experimental, the C++ source code will be made public after thorough debugging. The Java implementation is already available in the repository for running preliminary benchmarks. 
It is worth noting that tests conducted using ROOT version 6.32 showed faster read speeds for RNTuple, with an average time of  $1.87~seconds$, compared to $2.62~seconds$ for version 6.34. However, we were informed that version 6.34 incorporates format changes that will serve as the foundation for future developments, making it the version to be used moving forward.
More comprehensive performance tests are beyond the scope of this article and will be presented in a future publication.
The benchmarks are performed on an M1 Macbook Laptop with a 1 TB SSD drive, using JDK 21 for Java library and ROOT 6.34.08.

\begin{lstlisting}[language=c++, caption=Simplified version for filling histograms to compare data reading speeds more accurately., label=lst:simple_histo]
// declaration
int  h1i[100];int  h2i[100];int  h3i[100];int  h4i[100];
//.... LOOP over the entries
int b1 = (int) (ch1(i)*100);
int b2 = (int) (ch1(i)*100);
int b3 = (int) (ch1(i)*100);
int b4 = (int) (ch1(i)*100);
h1i[b1]++; h2i[b2]++;h3i[b3]++;h4i[b4]++;
//.... end of the LOOP
\end{lstlisting}

During our tests, we observed a performance difference between {\tt TH1D.Fill()} in ROOT and {\tt twig::h1d.fill()} (our custom implementation of a fast histogramming package), which impacted the overall benchmark results. To eliminate this variable, we replaced the histogramming code in both cases with a simplified version, as shown in Listing~\ref{lst:simple_histo}, and reran the benchmarks. The execution times obtained were 
$1.589$~seconds for ROOT (RNTuple) and $0.523$~seconds for HiPO, demonstrating that HiPO achieves read-and-fill times three times faster than ROOT RNTuple.

\section{Discussion}

The HiPO data format was developed specifically for CLAS12 to serve as a comprehensive solution for managing experimental data throughout its entire lifecycle. It is used by the CLAS12 reconstruction framework for data processing and production data storage. Over the past seven years, petabytes of data have been stored in HiPO, supporting the collaboration's physics analysis workflows with great success.

The adoption of a unified data format has enabled the development of standardized tools for common data manipulation tasks, such as data merging, filtering, and selective reduction. These tools eliminate the need for users to write and maintain custom code for routine operations. Additionally, graphical and text-based tools are available for browsing files and examining the content of individual events, which is particularly useful for debugging.

Recent advancements in columnar data storage within the HiPO format have introduced the capability to store data for physics analysis in a highly efficient manner. Tests have demonstrated that HiPO’s columnar storage delivers exceptional performance in data analysis, outperforming more established formats like ROOT and Parquet, which are widely used in High Energy and Nuclear Physics.

Future enhancements, including improved bindings for languages such as Python and Julia, are expected to further expand the usability of the HiPO format, making it an even more versatile tool for data analysis.

\section{Acknowledgments}

This material is based upon work supported by the U.S. Department of Energy, Office of Science,
Office of Nuclear Physics under contract DE-AC05-06OR23177, and NSF grant no. CCF-1439079 and
the Richard T. Cheng Endowment.


\end{document}